\documentstyle[ecssm]{article}

\newcommand{\beq}{\begin{equation}}
\newcommand{\eeq}{\end{equation}}
\newcommand{\bea}{\begin{eqnarray}}
\newcommand{\eea}{\end{eqnarray}}

\def\laq{~\raise 0.4ex\hbox{$<$}\kern -0.8em\lower 0.62
ex\hbox{$\sim$}~}
\def\gaq{~\raise 0.4ex\hbox{$>$}\kern -0.7em\lower 0.62
ex\hbox{$\sim$}~}

\def \ra {\rightarrow}

\def \Da {\Delta}
\def \b {\beta}
\def \a {\alpha}

\def \Ga {\Gamma}
\def \ga {\gamma}

\def \da {\delta}

\def \Om {\Omega}

\def \ls {\lambda_{\rm s}}

\begin{document}


\begin{flushright}

BA-TH/02-454\\
gr-qc/0301032
\end{flushright}

\vspace*{1.5 truein}

{\Large\bf\centering\ignorespaces
Phenomenology of the relic dilaton background
\vskip2.8pt}
{\dimen0=-\prevdepth \advance\dimen0 by23pt
\nointerlineskip \rm\centering
\vrule height\dimen0 width0pt\relax\ignorespaces

M. Gasperini 
\par}
\vspace{0.5 cm}
{\small\it\centering\ignorespaces
Dipartimento di Fisica , Universit\`a di Bari, 
Via G. Amendola 173, 70126 Bari, Italy\\
and \\
Istituto Nazionale di Fisica Nucleare, Sezione di Bari, Bari, Italy\\
\par}

\par
\bgroup
\leftskip=0.10753\textwidth \rightskip\leftskip
\dimen0=-\prevdepth \advance\dimen0 by17.5pt \nointerlineskip
\small\vrule width 0pt height\dimen0 \relax

\vspace*{1truein}

\centerline{\bf Abstract}

\noindent
We discuss the expected amplitude of a cosmic background of massive, 
non-relativistic dilatons, and we report recent results about its possible
detection. This paper  is a contracted version of a talk given at the 
15th SIGRAV Conference on ``General  Relativity and Gravitational
Physics" (Villa Mondragone, Roma, September 2002).

\vspace{2.5cm}
\begin{center}
------------------------------  

\vspace{0.5cm}
To appear in 
Proc. of the \\ 
{\sl ``15th SIGRAV Conference on General  Relativity and Gravitational
Physics"},\\
Villa Mondragone, Monte Porzio Catone (Roma), September 9-12, 2002\\
Eds. V. Gorini et al. (IOP Publishing, Bristol, 2003) 

\end{center}

\thispagestyle{plain}
\par\egroup

\vfill
\newpage


\setcounter{page}{1}

\title{Phenomenology of the relic dilaton background}

\author{Maurizio Gasperini\dag \ddag\footnote{E-mail:
gasperini@ba.infn.it}}

\affil{\dag\ Dipartimento di Fisica, 
Universit\`a di Bari, Via G. Amendola 173, 70126 Bari, Italy}

\affil{\ddag\ Istituto Nazionale di Fisica Nucleare, Sezione di Bari, Bari,
Italy}

\beginabstract
We discuss the expected amplitude of a cosmic background of massive, 
non-relativistic dilatons, and we report recent results about its possible
detection. 

\endabstract

\section{The amplitude of the dilaton background and the dilaton 
coupling-strength}

The aim of this talk is to  discuss the production and the possible
detection of a cosmic background of relic dilatons. The production is a
well known string cosmology effect  \cite{1}, so I will mainly
concentrate here on the interaction of the dilaton background with
a pair of realistic gravitational detectors \cite{2,3}. I will consider, in
particular, the case of {\em massive} and  {\em non-relativistic}
dilatons, where some new result has recently been obtained \cite{4}.

Let me start by recalling that the string effective action contains, 
already at lowest order, at least two fundamental fields, the metric and
the dilaton,  
\beq S= -{1\over 2 \ls^2} \int d^4x \sqrt{-g} e^{-\phi}
\left[ R+ \left(\nabla \phi\right)^2 + \dots\right]. 
\label{1}
\eeq
The dilaton $\phi$ controls the strength of all gauge interactions 
\cite{5} and, from a geometrical point of view, it may represent the
radius of the $11$-th dimension \cite{6} in the context of M-theory
and brane-world models of the Universe. What is important, for our
discussion, is that during the accelerated evolution of the Universe the
parametric amplification of the quantum fluctuations of the dilaton field
may lead to the formation of a stochastic background of relic scalar
waves \cite{1}, just like the amplification of (the tensor part of) metric
fluctuations may lead to the formation of a relic stochastic background
of gravitational waves \cite{7}. 

There are, however, two important differences between the graviton
and dilaton case. The first is that the dilaton fluctuations are
gravitationally coupled to the scalar part of the metric and matter
fluctuations. Such a coupling may lead to a final spectral distribution
different from that of gravitons. The second difference is that dilatons
(unlike gravitons) could be massive. Since the proper momentum is
redshifted with respect to the mass, then all modes tend to become
non-relativistic as time goes on, and a typical dilaton spectrum should
contain in general three branches: $\Om_1$ for relativistic modes,
$\Om_2$ for modes becoming non-relativistic inside the horizon, and 
$\Om_3$ for modes becoming non-relativistic outside the horizon. The
present energy-density of the background, per logarithmic
momentum-interval and in critical units, can thus be written as follows
\cite{1}: 
\bea 
&& \Om_1(p,t_0)=\left(H_1/ M_{\rm P}\right)^2
\Om_r(t_0)\left(p/ p_1\right)^{\da}, ~~~~~~~~~~~~~~~~~~~~~~
m<p<p_1, \label{2}\\ 
&& \Om_1(p,t_0)=\left(mH_1/  M_{\rm P}^2\right)
\left(H_1/ H_{\rm eq}\right)^{1/2} \left(p/ p_1\right)^{\da-1},
~~~~~~~~~p_m<p<m, \label{3}\\ 
&& \Om_1(p,t_0)= \left(m/ H_{\rm
eq}\right)^{1/2}\left(H_1/ M_{\rm P}\right)^2\left(p/ p_1\right)^{\da},
~~~~~~~~~~~~~~~~~~~~~~ p<p_m, 
\label{4} 
\eea
(see also \cite{3,8,9,10}). Here $m$ is the dilaton mass, $H_1$ the
inflation $\ra$ radiation transition scale, $H_{\rm eq}$ the radiation
$\ra$ matter transition scale,  $ \Om_r(t_0)$ the present radiation
energy-density, $\da$ the slope of the spectrum, $p_1$ the maximal
amplified momentum scale (i.e. the high-frequency cut-off parameter)
and $p_m=p_1(m/H_1)^{1/2}$ the transition scale corresponding to a
mode which becomes non-relativistic just at the time of horizon
crossing. If the mass is large enough, i.e. $m>p_1\sim (H_1/M_{\rm
P})^{1/2}10^{-4}$ eV, then all modes today are non-relativistic, and the
$\Om_1$ branch of the spectrum disappears. Note also that the
relativistic branch of the spectrum is typically growing \cite{1} in
minimal string cosmology models ($0<\da \leq 3$), while the
non-relativistic spectrum may have a flat or decreasing branch if $\da
\leq 1$. 

The present amplitude of the non-relativistic spectrum is  controlled by
two basic parameters, $H_1$ and $m$. In minimal string cosmology
models $H_1$ is typically fixed at the string scale, $H_1 \simeq M_{\rm
s}$, and the question {\em ``How strong is today the relic dilaton
background?"}  thus becomes {\em ``How large is the dilaton mass?"} 
We would like to have a mass small enough to avoid the decay and to
resonate with the present gravitational antennas, but large enough to
correspond to a detectable amplitude. It is possible?

The answer depends on the value of the dilaton mass.  From the
theoretical side we have no compelling prediction, at present. From the
phenomenological side, however, we know that the allowed values of
mass (and thus of the range of the dilatonic forces) are strictly
correlated to the strength of the dilaton coupling to macroscopic
matter. We are thus lead to the related question:  {\em ``How large is
the dilaton coupling?"}

For a better formulation of such a question we should recall that the
motion of a macroscopic body, in a gravi-dilaton background, is in
general non-geodesic, because of the direct coupling to the gradients
of the dilaton field \cite{2}, 
\beq
{du^\mu\over d\tau}+ \Ga_{\a\b}^\mu u^\a u^\b+q \nabla^\mu \phi=0,
\label{5}
\eeq
induced by the effective ``dilatonic charge" $q$ of the body, which
controls the relative strength of scalar-to-tensor forces. So, how large
is $q$? 

The experimental value of $q$, as is well known, is directly constrained
by tests of the equivalence principle and of macroscopic  gravity, 
which provide exclusion plots in the $\{m, q^2\}$ plane
(see for instance \cite{11}). From a theoretical point of view, on the
other hand, there are two possible (alternative) scenarios. A first
possibility is that, by including all required loop corrections into the
effective action, the dilatonic charge $q$ becomes large and
composition-dependent \cite{12}. In that case one has to impose $m
\gaq 10^{-4}$ eV, to avoid contradiction with the present gravitational
experiments, which exclude testable deviations from the standard
Newton's law down to the millimeter scale \cite{13}. The alternative
possibility is that, when including loop corrections to all orders, the
resulting effective charge of ordinary macroscopic matter turns out to
be very small ($q \ll1$)and universal \cite{14}. In that case the dilaton
mass can be very small, or even zero. In the following discussion we
will accept this second (phenomenologically more interesting)
possibility, relaxing however the assumption of universal dilaton
interactions (which is not very natural, and which seems to require
fine-tuning). 

We shall thus assume that the relic dilatons are weakly coupled to
matter, and light enough to have not yet decayed ($m \laq 100$ MeV),
so as to be available to present observations. The amplitude of the
background, as already noticed, depends however on $m$, and a larger
mass corresponds in principle to a stronger signal. In this range of
parameters, on the other hand, there is an upper limit on the mass
following from the critical density bound, 
\beq
\int^{p_1} d(\ln p) \Om(p) \laq 1,
\label{6}
\eeq
which has to be imposed to avoid a Universe over-dominated  by
dilatons. So, which values of mass may correspond to the stronger
signal, i.e. to an (almost) critical dilaton background?

The answer depends on the shape of the spectrum. 
If $m>p_1$ and $\da \geq 1$, so that all modes are non-relativistic and
the spectrum is peaked at $p=p_1$, the critical bound reduces to
$\Om_2(p_1)\laq 1$, and implies 
\beq 
m<\left(H_{\rm eq}M_{\rm
P}^4/H_1^3\right)^{1/2}, \label{7}
\eeq
which, for $H_1 \sim M_{\rm s}\sim 10^{-1} M_{\rm P}$, is saturated by
$m \sim 10^2$ eV. This means that the density of the relic dilaton
background may approach the critical upper limit for values of mass
which are too large (as we shall see later) to fall within the resonant
band of present gravitational detectors. If $\da<1$ the above bound is
relaxed, but the mass is still constrained to be in the range $m>p_1$. 

Let us thus
consider the case $m<p_1$, assuming that the spectrum is flat enough
($\da <1$) to be dominated by the non-relativistic branch, peaked at
$p=p_m$. The critical density bound then reduces to $\Om_2(p_m)\laq
1$, and implies
\beq
m<\left(H_{\rm eq}M_{\rm P}^4 H_1^{\da-4}\right)^{1/(\da+1)}.
\label{8}
\eeq
For $H_1=M_{\rm s}$ and $\da \ra 0$ this bound is saturated by
masses as small as $m \sim 10^{-23}$ eV, and this opens an interesting
phenomenological possibility. Indeed, if the mass is small enough to fit
the sensitivity range of present antennas (see the next section), the
weakness of the dilaton coupling ($q \ll1$) could be compensated by a
large (i.e., near to critical ) background intensity, much larger than for
gravitons:
\beq
\Om_{\rm grav} \laq 10^{-6} \ll \Om_{\rm dil} \laq 1.
\label{9}
\eeq
 Massless backgrounds (like gravitons) are indeed constrained by the
nucleosynthesis bound \cite{15}, which may be evaded by massive
backgrounds dominated by the non-relativistic branch of the spectrum.

\section{Detection of a cosmic background of ultra-light (but
non-relativistic) scalar particles}

Starting from the equation of motion (\ref{5}) it  can be easily deduced
that the full coupling of dilatons to a gravitational detector is described
by a generalized equation of geodesic deviation \cite{2,3}: 
\beq 
{D^2\eta^\mu\over D\tau^2} +R_{\b\a\nu}~^\mu 
u^\a u^\nu \eta^\b + q
\eta^\b\nabla_\b \nabla^\mu \phi=0. \label{10} 
\eeq
A relic dilaton background can thus interact with  a gravitational
antennas in two ways:

(i) {\em indirectly}, through the geodesic  coupling of the gravitational
charge of the detector to the scalar part of the metric fluctuations
induced by the dilatons \cite{16};

(ii) {\em directly},  through the non-geodesic coupling of the scalar 
charge of the detector to the gradients of the dilaton background
\cite{2,3}.

The indirect coupling has gravitational  strength ($q=1$), but the
amplitude of the gravitational background is expected to be (in general)
highly suppressed ($\Om \ll1$); the direct coupling may refer to a much
higher amplitude ($\Om \laq1$), but the effective charge $q$ has
to be strongly suppressed to agree with the results of the present
gravitational experiments (see \cite{2,3}). 

Taking into account both possibilities, the detection of a stochastic
background of relic dilatons, with generic spectrum $\Om(p)$, is
controlled by the ``signal-to-noise" ratio SNR, determined by the
cross-correlation of the outputs of two antennas \cite{3}:
\beq
{\rm SNR}={H_0^2\over 5 \pi^2}\left[2T \int_0^\infty 
{dp\over p^3 (m^2+p^2)^{3/2}}
{\Om^2(p) \ga^2(p)\over S_1(\sqrt{m^2+p^2}) 
S_2(\sqrt{m^2+p^2})}\right]^{1/2}.
\label{11}
\eeq
Here $S_1(f)$ and $S_2(f)$ are the noise power-spectrum of the two
detectors, $T$ the observation time, $\ga(p)$ the ``overlap function" in
momentum space, 
\beq
\ga(p)= {15\over 4\pi}\int {d^2 \hat n } F_1(\hat n) F_2 (\hat n) 
 e^{2\pi i p \hat n \cdot ({\vec x}_2 - {\vec x}_1)},
~~~~~~~
F(\hat n)=q e_ab(\hat n) D^{ab}.
\label{12}
\eeq
$F_1$ and $F_2$ are the ``pattern functions" of the  two detectors,
$e_{ab}$ the polarization of the scalar wave propagating along the
direction specified by the unit vector $\hat n$, and $D^{ab}$ the
``response tensor" taking into account the specific geometry and
orientation of the arms of the detectors. The ratio (\ref{11}) differs
from a similar expression for gravitons \cite{17} because of the mass,
of the different polarization of the scalar wave, and of the pattern
function which is in general proportional to the effective charge of the
detector. 

The presence of the mass, in particular, is crucial for  determining the
(possible) resonant response of the detectors, whose noise spectrum
has a minimum inside a frequency band $f_0$ (typically, $f_0 \sim
10-10^3$ Hz for present detectors). Outside this sensitivity band the
noise diverges, and the signal is negligible. Since $f=\sqrt{m^2+p^2}$,
we have three possibilities. 

If $m \gg f_0$ then the noise is always 
outside the sensitivity band, as $S(f) \gg S(f_0)$ for all modes $p$ of the
spectrum. If $m \sim f_0$ the noise, on the contrary, is within the
sensitivity band for all non-relativistic modes, as $S(f) \simeq S(m)
\sim S(f_0)$ for  $p \leq m$. Finally, if $m \ll f_0$, then the sensitivity
band may (possibly) overlap only with the relativistic branch of the
spectrum, when $p= f \simeq f_0$. Taking into account these various
possibilties, the analysis of the SNR induced in a pair of interferometric
and/or resonant detectors leads to the following results.

1) A resonant response  to a {\em massive},  {\em
non-relativistic} background of scalar particles is possible, provided the
mass is in the sensitivity band of the detector \cite{3}. For the
present resonant frequencies, $f_0 \sim 1$ Hz -- $1$ kHz, 
the present sensitivity is thus in the mass-range   
\beq 
10^{-15} {\rm eV} \laq m \laq 10^{-12}
{\rm eV}. 
\label{13} 
\eeq

2) For the differential mode of the interferometers, whose  response
tensor is traceless ($D^{ab}\da_{ab}=0$), the non-relativistic overlap
function is always proportional to the relativistic one \cite{3}, 
\beq
\ga_{\rm non-rel}(p)=\left[p^4/ (m^2+p^2)^2\right]\ga_{\rm rel}(p),
\label{14} 
\eeq
both for the direct  and indirect coupling. This proportionality
factor induces a strong suppression for the SNR of non-relativistic
modes, with $p \ll m$; as a consequence, the detection is in principle
allowed (with the planned sensitivities of second-generation detectors)
only if the spectrum $\Om(p)$ is peaked at $p=m\sim f_0$ \cite{3}.

3) The above suppression, however, may be absent in the case of
resonant mass detectors, and in particular for the geodesic (indirect)
coupling of the dilaton background to the monopole mode of a sphere
(whose  response tensor is trivial, $D^{ab}=\da^{ab}$). For two spheres,
with spatial separation $d$, we have indeed \cite{3,4} 
\beq 
\ga (p)={15\over 2\pi}\left(3 m^2 +2 p^2\over m^2
+p^2\right)^2 {\sin (2\pi p d)\over pd}, ~~~~~
\ga_{\rm non-rel}(p \ra 0)={9\over 4}\ga_{\rm
rel}(m  \ra 0). 
\label{15}
\eeq

3) The response of spherical detectors to a massive stochastic 
background is also particularly enhanced (with respect to
interferometers) for a flat enough spectrum \cite{4}. For instance, if 
$\Om \sim p^\da$, and $\da <1$, then the  SNR grows with the
observation time like $T^{1-\da/2}$, i.e. faster than the usual $T^{1/2}$
dependence. 

In order to illustrate this final, important result, let us consider a
non-relativistic, growing spectrum with slope $\da <1$, peaked at $p
\sim m \sim f_0$. For $p \ra 0$ the noise and the overlap functions of
the two correlated monopole modes go to a constant, and the SNR
integral (\ref{11}) would seem to diverge, if extended down to the
lower limit $p=0$. The SNR integral, however, has an infrared cut-off
at a finite value $p_{\rm min}$ determined by the {\em minimum
resoluble frequency-interval} $\Da f$ associated to $T$, and defined by 
$ \Da f= (p^2+m^2)^{1/2}-m \geq T^{-1}$, 
 which gives $p_{\rm min} \simeq (2m/T)^{1/2}$. When $\da <1$ the SNR
integral is dominated by its lower limit, and this introduces the
anomalous $T$-dependence of the signal,
\beq
{\rm SNR} \sim T^{1/2}\left[\int^m_{p_{\rm min}}dp p^{2\da-3}
\right]^{1/2}\sim 
T^{1-\da/2}, ~~~~~ \da \leq 1
\label{16}
\eeq
(see \cite{4} for numerical examples). 

\section{Conclusion}
 
The possible production of a cosmic background of relic dilatons is a 
peculiar aspect of string cosmology and of the pre-big bang scenario
\cite{1,10}. Light dilatons are particularly interesting, in this context,
because if $m \laq 100$ Mev the dilatons of the background have not
yet decayed, if $m \laq 10$ keV they could provide a significant
contribution to the present cold dark-matter density, and if $m \laq
10^{-12}$ eV they are also in principle detectable \cite{2,3,4} by
``advanced" (i.e., second-generation) interferometers and by (future)
resonant spheres (provide the mass is in the sensitivity band of the
antennas). Spherical detectors, in particular, seem to be favoured with
respect to interferometers if the non-relativistic dilaton spectrum is
flat enough, and  if it {\em is not} peaked at $p=m$. In any case, the
search for cosmic dilatons (or, more generally, for relic scalar particles)
may provide unique infomation on primordial cosmology and Planck-scale
physics. 

\section*{Acknowledgements}

It is a pleasure to thank E. Coccia, C. Ungarelli and G. Veneziano 
for useful discussions and for fruitful collaboration on the
production and detection of a relic dilaton background.

\end{document}